\documentclass[twocolumn,preprintnumbers,secnumarabic,amsmath,amssymb,superscriptaddress,nofootinbib]{revtex4}

\usepackage{graphicx}
\usepackage{xcolor}
\usepackage{dcolumn}
\usepackage{bm}
\definecolor{linkcolor}{rgb}{0.0,0.3,0.5}
\usepackage[unicode, colorlinks=true, linkcolor=linkcolor, citecolor=linkcolor, 
filecolor=linkcolor,urlcolor=linkcolor, pdfusetitle]{hyperref}

\usepackage{makecell}
\usepackage{nccmath}

\usepackage{stackengine}

\newcommand{\beq}{\begin{equation}}
\newcommand{\eeq}{\end{equation}}

\newcommand{\bea}{\begin{eqnarray}}
\newcommand{\eea}{\end{eqnarray}}

\def\IR{\mathbb{R}}

\begin{document}

\vspace*{-.1cm}
\title{The Uplifton}

\author{Iosif Bena}
\affiliation{{Institut de Physique Th\'eorique,
	Universit\'e Paris Saclay, CEA, CNRS,
 F-91191 Gif-sur-Yvette, France
}}
\author{Pierre Heidmann}

\affiliation{{Department of Physics and Astronomy,
Johns Hopkins University, Baltimore, MD 21218, USA}}

\preprint{}

\begin{abstract}

Almost all proposals to construct de Sitter vacua with a small cosmological constant involve flux compactifications with stabilized moduli. These give AdS vacua, which are uplifted to de Sitter by adding antibranes in certain regions of the compactification manifold. However, antibranes are charged, singular and interact nontrivially with other ingredients of the compactification; this can invalidate the de Sitter construction. In this Letter,  we construct a new ingredient for uplifting AdS solutions to de Sitter,  which is neutral, smooth and horizonless, and therefore bypasses some of the problems of antibrane uplift.


\end{abstract}

\maketitle

\section{Introduction}
\vspace*{-.3cm}

The accelerated expansion of our Universe points to the existence of a positive vacuum energy. However, String Theory appears rather reluctant to provide four-dimensional solutions with a positive vacuum energy. Even the simplest solution with positive vacuum energy -- de Sitter space -- is very hard to construct. There are no-go theorems preventing the direct realization of such a space via a compactification with common ingredients \cite{Maldacena:2000mw}. The most popular scenario to bypass these theorems and construct de Sitter spaces with a small cosmological constant, proposed by Kachru, Kallosh, Linde and Trivedi \cite{Kachru:2003aw}, involves a rather intricate sequence of steps: one first stabilizes the complex-structure moduli by turning on topologically-nontrivial fluxes on the compactification manifold and then one stabilizes the K\"ahler moduli via nonperturbative effects. This results in an Anti de Sitter solution (with a negative cosmological constant) which should be ``uplifted'' to a de Sitter solution (with a positive cosmological constant) by placing D3 branes with negative charge (antibranes) in a region of large warping inside the compactification manifold. 

This last step has been under a lot of intense scrutiny over the past fifteen years, because the $\overline{\text{D3}}$ branes interact non-trivially with the fluxes used to stabilize the moduli \cite{Bena:2009xk}. This can result in tachyons, unexpected massless modes and runaways \cite{Bena:2018fqc,Bena:2014jaa}. Furthermore, the interaction of the four-form field sourced by the antibrane with the other fields of the compactification can give rise to new flux components, that can affect the regime of validity of K\"ahler moduli stabilization \cite{Bena:2022ive}.  

The purpose of this Letter is to construct a new ingredient for uplifting the cosmological constant - an uplifton - which is neutral, smooth and horizonless, thereby avoiding some of the problems of antibranes. Furthermore, unlike antibranes  \cite{Aharony:2005ez}, the uplifton cannot move around the compactification manifold . 

At first glance, neutral solutions which have mass but no charge do not appear to be optimal ingredients for use in flux compactifications, because they na\"ively source a metric that does not preserve Lorentz invariance in the spacetime directions. The best example is perhaps a non-extremal black D3 brane, which does have more mass than charge, but whose metric breaks the Lorentz symmetry along the spacetime direction.\footnote{The time-time and the parallel space-space components of this metric are different functions of the radius.} However, as we will see, the non-extremal solutions we construct preserve this Lorentz invariance through a novel mechanism which involves the shrinking of certain compact directions.

To construct these solutions,  we use the formalism that has been developed over the past few years by one of the authors and Bah  \cite{Bah:2020pdz,Bah:2021owp,Bah:2021rki,Heidmann:2021cms,Bah:2022yji,Bah:2022pdn}: the equations governing certain supergravity solutions with $D-2$ commuting Killing vectors and suitable fluxes decompose into a set of Ernst equations, thereby admitting an integrable structure. This formalism has allowed to obtain a plethora of solutions, describing both bound states of bubbles and black holes, as well as smooth horizonless solutions with multiple bubbles and topologically non-trivial fluxes. Some of these solutions are non-extremal and charged, but it is also possible to construct neutral solutions with opposite fluxes wrapping different cycles \cite{Bah:2022yji}.

At first glance, the simplest way to construct an uplifton appears to be using smooth bubbles with D3-brane and $\overline{\text{D3}}$ brane charges. However, the mechanism by which the solutions of  \cite{Bah:2022yji} carry charges involves topologically-nontrival cycles formed by the shrinking of at least one direction inside the brane worldvolume. Hence, bubbling solutions whose bubbles have D3 and $\overline{\text{D3}}$ charges break the $SO(3,1)$ Lorentz invariance.

Since we are looking for upliftons that one can add to Type IIB flux compactifications, the obvious step to bypass this problem is to use the technique of  \cite{Bah:2022yji}  to construct neutral solutions with D5 and $\overline{\text{D5}}$ bubbles that preserve the $SO(3,1)$ invariance.

In this Letter we present the simplest of these solutions and their use in flux compactifications, leaving the details of their construction to a companion paper. Even if the upliftons  are neutral, they have a nontrivial magnetic three-form field strength profile which is both positive and negative, so locally they have  D5 and $\overline{\text{D5}}$ charge corresponding to branes extending along $(t,x_1,x_2,x_3,x_4,x_5)$, where $x_5$ must be compact. The orthogonal space is a $U(1)$ fibration of a compact coordinate, $y$, over a three-dimensional base given in spherical coordinates $(r,\theta,\phi)$.  In a flux compactification, both $x_4$ and $x_5$, together with $y, r,\theta,\phi$ will be part of the compactification manifold. As we will show, regularity will impose certain periodicity constraints on the $x_5$ and $y$ coordinates, so, unlike $\overline{\text{D3}}$ branes, the uplifton will not be able to move inside the compactification manifold. 

\vspace*{-.5cm}
\section{The uplifton 
solution}
\label{sec:solution}
\vspace*{-.2cm}

\begin{figure}
\begin{center}
\includegraphics[width= 0.5 \textwidth]{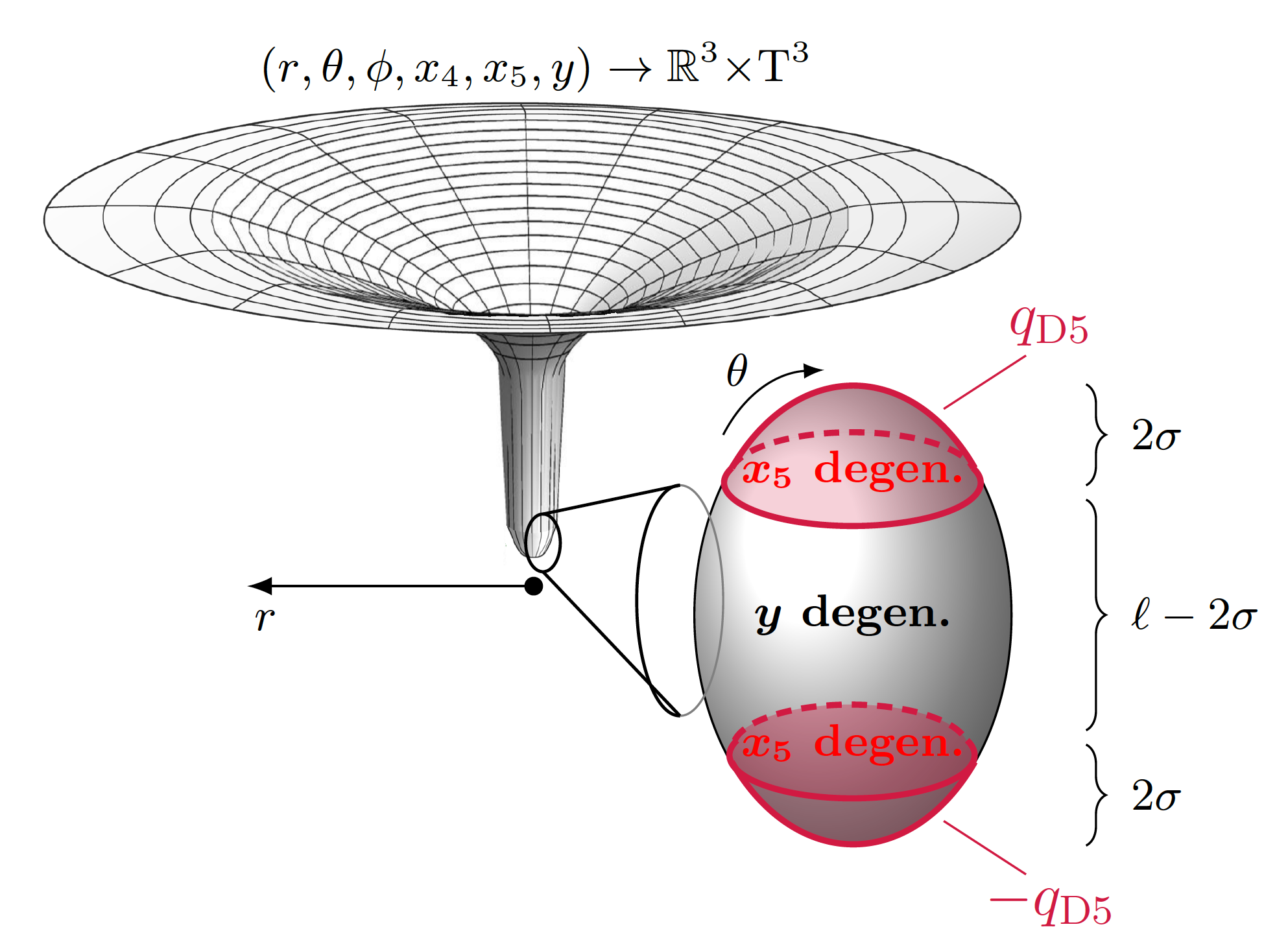}
\caption{Schematic description of the uplifton internal directions.  The spacetime smoothly terminates at $r=\ell+2\sigma$ where the $y$ and $x_5$ circles degenerate alternatively as smooth bolts.  The $x_5$-bolts carry opposite D5 brane charges.}
\label{fig:SolProfile}
\end{center}
\end{figure}

The six-dimensional spacetime transverse to the directions of the uplifton, $(t, x_1,x_2,x_3)$,  is made from three compact circles fibered over a three-dimensional space that becomes asymptotically $\IR^3$. Using the technology of \cite{Bah:2022yji} we can build upliftons with an arbitrary number of bubbles, but for simplicity we will present here the simplest upliftons.  They have two identical bubbles carrying opposite D5 charges, and located at the opposite ends of an uncharged bubble (see Fig.\ref{fig:SolProfile}).  The solutions are determined in terms of three parameters $(\ell,m,q)$: $\ell$ and $m$ are related to the mass, or energy, induced by the sources,  while $q$ is related to the amplitude of the D5 charges carried by the outermost bubbles.

Being a bound state of three sources,  we introduce three sets of local spherical coordinates,  centered around each bubble:
\begin{align}
4 r_1 &\,\equiv\, r_-^{(0)} +r_-^{(1)}-4 \sigma\,, \quad \cos \theta_1 \,\equiv\,  \frac{r_-^{(0)} -r_-^{(1)}}{4\sigma}\,,\nonumber \\
4 r_2 &\,\equiv\, r_-^{(1)} +r_+^{(1)}-2(\ell-2 \sigma)\,, \quad \cos \theta_2 \,\equiv\,  \frac{r_-^{(1)} -r_+^{(1)}}{2(\ell-2 \sigma)}\,,\nonumber \\
4 r_3 &\,\equiv\, r_+^{(1)} +r_+^{(0)}-4 \sigma\,, \quad \cos \theta_3 \,\equiv\,  \frac{r_+^{(1)} -r_+^{(0)}}{4\sigma}\,, 
\label{rdef}
\end{align}
where we have defined $2\sigma$ and $\ell-2\sigma$ to be the size of the outermost bubbles and the middle bubble respectively,  and the distance $(r_\pm^{(0)},r_\pm^{(1)})$,  related to the $\IR^3$ spherical coordinates $(r,\theta)$ such as
\begin{align}
\sigma & \,\equiv \, \sqrt{m^2-q(q-2\gamma)}\,,\qquad \gamma \,\equiv \, \frac{2mq}{\ell+2m}\,,\nonumber \\
r_\pm^{(0)}& \equiv 2 r - (\ell+2\sigma)(1\pm \cos \theta)\,,  \\
r_\pm^{(1)}  &\equiv\medmath{\sqrt{((2(r-\sigma)-\ell)\cos \theta\pm(2\sigma-\ell))^2+4 r (r-\ell-2\sigma)\sin^2\theta}} .\nonumber
\end{align}

These solutions exist when the parameters satisfy \cite{Bah:2022yji}:
\begin{equation}
\ell > 2 m \,,\qquad q < m \,\sqrt{\frac{\ell+2m}{\ell-2m}}.
\label{eq:validitybound}
\end{equation}
When the second bound is saturated, the outer bubbles degenerate into two singular five-brane sources of opposite charges. Hence, in this extreme regime, the solution can be thought of point-like D5 and $\overline{\text{D5}}$ branes at the opposite ends of a bolt.

\begin{widetext}
The string-frame type IIB uplifton solution is given by
\begin{align}
ds_{10}^2 \,=\, &\frac{1}{\sqrt{Z}}\left[- dt^2+ dx_1^2+ dx_2^2+dx_3^2+dx_4^2+\frac{r_1 \,r_3}{(r_1+2\sigma)(r_3+2\sigma)} \,dx_5^2 \right] \nonumber \\
& \,+\, \sqrt{Z}\left[ f \left(\frac{dr^2}{1-\frac{\ell+2\sigma}{r}} +r^2 \,d\theta^2 \right) +r^2 \sin^2\theta d\phi^2\,+\, \frac{r_2}{r_2+\ell-2\sigma} dy^2\right] \label{eq:Met} \\
C^{(2)}\,=\,& H d\phi\wedge dy\,,\qquad e^\Phi \,=\, \frac{1}{\sqrt{Z}}\,,\qquad B_2 \,=\, C^{(0)} \,=\, C^{(4)} \,=\, 0\,.\nonumber
\end{align}
where the definitions of $r_1,r_2,r_3$ are in equation \eqref{rdef}, and we have introduced the following warp factors and gauge potentials
\begin{align}
Z\equiv & \frac{(r_1+\sigma + m)(r_3+\sigma +m)+\left(q-\gamma(1+\cos \theta_3)\right)\left(q-\gamma(1-\cos \theta_1)\right)}{\sqrt{\left((r_1+2\sigma)^2+\gamma^2\,\sin^2 \theta_3\,\left(1+\frac{2\sigma}{r_1}\right)\right)\left((r_3+2\sigma)^2+\gamma^2\,\sin^2 \theta_1\,\left(1+\frac{2\sigma}{r_3}\right)\right)}},  \hspace{7cm} \nonumber
\end{align}
\begin{align}
f^4\equiv & \frac{(r_1(r_1+2\sigma)+\gamma^2\sin^2\theta_3)(r_3(r_3+2\sigma)+\gamma^2\sin^2\theta_1)}{(1+2\delta)^2\,\left(r_1+\sigma(1-\cos \theta_1) \right)^4\left(r_3+\sigma(1+\cos \theta_3) \right)^4}\left(1+2\delta \,\frac{(q-\gamma)(r_1-r_3)+(\gamma m-\ell(q-\gamma))(\cos \theta_1+\cos \theta_3)}{(q-\gamma)(r_3-r_1)+\gamma m (\cos \theta_1+\cos \theta_3)}\right)^2 \nonumber\\ 
&\times  \frac{r_1^2 r_2^4 r_3^2 \,(r_2+\ell-2\sigma)^2(r-\ell-2 \sigma)^{-2}}{\left(\left(r_2+\frac{\ell}{2}-\sigma(1-\cos \theta_1) \right)^2-\frac{\ell^2}{4}\right)\,\left(\left(r_2+\frac{\ell}{2}-\sigma(1+\cos \theta_3) \right)^2-\frac{\ell^2}{4}\right)}\,,\qquad \delta \equiv \frac{m^2(\ell+2m)^2+\ell^2 q^2}{(\ell+2m)^2(\ell^2-2m^2)+2\ell^2 q^2}\,,\nonumber \\
H \,=\, & \frac{(q-\gamma)(r_3-r_1)+\gamma m (\cos \theta_1+\cos\theta_3)}{(r_1+\sigma + m)(r_3+\sigma +m)+\left(q-\gamma(1+\cos \theta_3)\right)\left(q-\gamma(1-\cos \theta_1)\right)} \Biggl[ \frac{(r_1+\sigma)\cos\theta_1+(r_3+\sigma) \cos \theta_3}{2}  \\
& \hspace{5.5cm} -\frac{\ell}{2} \left(1+\frac{\gamma^2}{m} \right) \,\frac{(r_3+\sigma+m)(2m\cos\theta_1+\ell)-(r_1+\sigma+m)(2m\cos\theta_3-\ell)}{\ell(r_1-r_3)+2(q-\gamma)^2 (\cos \theta_1+\cos \theta_3)} \Biggr] \nonumber \\
& +2 m(q-\gamma) \frac{2(r_1-r_3)-\ell(\cos \theta_1+\cos \theta_3)}{\ell(r_1-r_3)+2(q-\gamma)^2 (\cos \theta_1+\cos \theta_3)}\,. \nonumber
\end{align}
\end{widetext}
The spacetime is smooth and terminates at $r=\ell+2\sigma$. At this locus, either $r_1$ or $r_2$ or $r_3$ is zero, depending on the value of $\theta$ in terms of three intervals. These intervals are determined by the critical angle, $\theta_c$, defined as:
\begin{equation}
\cos \theta_c\, \equiv\, \frac{\ell-2\sigma}{\ell+2\sigma}\,.
\end{equation} 
For $0\leq\theta\leq \theta_c$ and $\pi-\theta_c\leq \theta\leq \pi$,  $r_3=0$ and $r_1=0$ respectively,  such that $x_5$  degenerates at the origin.  For $\theta_c\leq \theta\leq \pi-\theta_c$,  $r_2=0$ and the $y$ coordinate degenerates. As we will see in the next section, these coordinate degeneracies correspond to smooth bolts only if $x_5$ and $y$ are compact.\footnote{This is a general feature of solutions constructed using the procedure of \cite{Bah:2020pdz,Bah:2021owp,Bah:2021rki,Heidmann:2021cms,Bah:2022yji,Bah:2022pdn}, and the reason why one cannot construct a bound state with D3 and $\overline{\text{D3}}$ charges without having to compactify one of the internal D3 directions and break the $SO(3,1)$ Lorentz invariance.} Thus, the maximal Lorentz invariance our solutions can preserve is $SO(4,1)$, when $(x_1,x_2,x_3,x_4)$ are infinite. However,  we can also compactify one of these directions ($x_4$ for example) to obtain a more general solution which only preserves $SO(3,1)$ and which can be embedded in a flux compactification.

\vspace*{-.5cm}
\section{The gluing properties}
\label{sec:Reg}
\vspace*{-.2cm}

To analyse the regularity of the uplifton,  it is useful to denote the periodicities of $(y, x_4,x_5)$ as $R_y, R_{x_4}$ and $R_{x_5}$:
\begin{equation}
y= y + 2\pi R_y\,,\quad x_4=x_4 + 2\pi R_{x_4} \,,\quad x_5=x_5 + 2\pi R_{x_5}\,.
\end{equation}

As we explained above, the shrinking of the $y$ and $x_5$ coordinates at the origin ($r=\ell+2\sigma$) depends on the three $\theta$-intervals   $0\leq\theta_c \leq \pi-\theta_c\leq \pi$. The local coordinates  adapted for the first, second and third interval respectively, are $(r_3,\theta_3)$,  $(r_2,\theta_2)$ and $(r_1,\theta_1)$ \eqref{rdef}. They allow us to write the constant-time slices of the metrics when each $r_i\to0$ at the origin as
\begin{equation}
ds_{10}^2 \bigl|_{dt=0}\,  \propto \,  \frac{dr_i^2}{r_i}+\frac{4 r_i}{C_i}\,dX^2 + ds(\mathcal{K}_7)^2\,,
\end{equation}
where $C_i$ are constants that depends on $(\ell,m,q)$. The periodic coordinate, $X$ stands for $x_5$ when $i=1,3$ and for $y$ when $i=2$,  while $\mathcal{K}_7$ describes a smooth orthogonal space of topology S$^3\times$S$^1\times\mathbb{R}^3$ when $i=1,3$ and  S$^2\times$T$^2\times\mathbb{R}^3$ when $i=2$.  The bolt structure is explicit in terms of the radial coordinates $\rho_i^2\equiv4 r_i$ and the $\mathbb{R}^2$ has no conical singularity if $R_X^2=C_i$.  This requires:
\begin{equation}
\begin{split}
R_y^2& = \frac{4(\ell^2-4m^2)(\ell^2-4\sigma^2)}{\ell^2}\,,\\
 R_{x_5}^2&=\frac{4(\ell+2m)(\ell+2\sigma)(m+\sigma)}{\ell}\,.
 \end{split}
\end{equation}
Moreover,  the three-form field strength is regular everywhere, and its integrals on the first and third bolts are equal and opposite. They give the D5 and $\overline{\text{D5}}$ quantized charges carried by these bolts:
\begin{equation}
\begin{split}
N_{\text{D5}}&\,=\,\frac{1}{4\pi^2 g_s l_s^2}\,\int_{\theta_1 \phi y} F_3\bigl|_{r_1=0} \,=\,  \frac{2R_y \,q}{g_s l_s^2}\,,\\
N_{\overline{\text{D5}}}&\,=\,-\frac{1}{4\pi^2 g_s l_s^2}\,\int_{\theta_3 \phi y} F_3\bigl|_{r_3=0} \,=\,  \frac{2R_y \,q}{g_s l_s^2}\,,
 \end{split}
\end{equation}
where $g_s$ is the string coupling and $l_s$ is the string length.  One can absorb the $\sqrt{g_s}\,l_s$ coefficient by expressing all length scales in units of $\sqrt{g_s} \, l_s$. This is done by rescaling 
\begin{equation}
(\ell,m,q,\sigma,R_X)\, \equiv\, \sqrt{g_s} \,l_s \times (\bar{\ell},\bar{m},\bar{q},\bar{\sigma},\bar{R}_X).
\label{rescaling}
\end{equation}

Our solutions have three parameters, and two regularity constraints; it is natural to choose the free parameter to be the quantized charge of the bubbles, $N=N_{\text{D5}}=N_{\overline{\text{D5}}}=2 \bar{R}_y \bar{q}$. The parameters of the regular solutions are therefore completely determined by the periodicities of $y$ and $x_5$ at infinity and by the D5 quantized charge:
\begin{equation}
\begin{split}
\bar{m}& \,=\,  \frac{1}{4 \bar{R}_{x_5}} \left( \frac{ \bar{R}_{x_5}^3}{\sqrt{4  \bar{R}_{x_5}^2+ \bar{R}_y^2}}+ \frac{N^2}{\sqrt{4 N^2+ \bar{R}_{x_5}^2  \bar{R}_y^2}} \right)\,, \\
\bar{\ell} & \,=\, 2\bar{m} +\frac{ \bar{R}_y^2}{4 } \left( \frac{ 1}{\sqrt{4  \bar{R}_{x_5}^2+ \bar{R}_y^2}}+ \frac{\bar{R}_{x_5}}{\sqrt{4 N^2+ \bar{R}_{x_5}^2  \bar{R}_y^2}}\right),\\
\bar{\sigma} &\,=\, \frac{1}{4  \bar{R}_{x_5}} \left( \frac{ \bar{R}_{x_5}^3}{\sqrt{4  \bar{R}_{x_5}^2+ \bar{R}_y^2}}- \frac{N^2}{\sqrt{4 N^2+ \bar{R}_{x_5}^2  \bar{R}_y^2}}  \right),
\label{eq:lmsigma}
\end{split}
\end{equation}
and the validity bound \eqref{eq:validitybound} translates into
\begin{equation}
0 \,\leq\, N \,<\,  \bar{R}_{x_5}^2. \label{constraint}
\end{equation}
The solutions have two simple limits. When $N=0$,  the flux of the solution is strictly zero, and the solution becomes a pure-gravity solution describing three colinear vacuum bolts. When $N= \bar{R}_{x_5}^2$ the size of the outwards bubbles becomes zero ($\sigma=0$) and these bubbles degenerate to singular locally-supersymmetric D5 and $\overline{\text{D5}}$ branes on a vacuum bubble.

\vspace*{-.5cm}
\section{Uplifting with the uplifton}
\label{sec:uplifting}
\vspace*{-.2cm}

In order to use the uplifton for uplifting  we have to  embed it in flux compactifications, and  compare its energy with that of other uplifting ingredients, such as five-branes or anti-D3 branes. Our solution has three compact circles and one can consider adding it to a region of the compactification manifold where the geometry looks locally like a $U(1)^3$ fibration over $\IR^3$. The mass of this solution is then completely determined by the size of the three $U(1)$'s in this region and by the quantized five-brane and anti-five-brane charges of the bubbles.

To compute the mass,  we reduce the uplifton along $x_{1,2,3,4,5}, y$, and obtain a geometry with $\IR^{3,1}$ asymptotics:\footnote{It is also possible to construct upliftons in which $y$ is fibered over the $\mathbb{R}^3$ base and the asymptotics is $\IR^{4,1}$.}
\begin{equation}
ds_4^2 \,=\, - \sqrt{\frac{1-\frac{\ell+2\sigma}{r}}{Z}}\,dt^2 + \sqrt{Z\,\left(1-\frac{\ell+2\sigma}{r}\right)}\,ds_3^2\,,
\end{equation}
where $ds_3^2$ is the three-dimensional base in the bracket of \eqref{eq:Met}.  The ADM mass per unit of spacetime volume (parameterized by $x_1,x_2$ and $x_3$) is\footnote{The relation between the four-dimensional and ten-dimensional Newton constants is  $G_{10}=8\pi^6 g_s^2l_s^8=(2\pi \sqrt{g_s}l_s)^3 \bar{R}_{x_4}\bar{R}_{x_5}\bar{R}_y\, G_4 $.}
\begin{align}
\mathcal{M} &\,=\,  \frac{\ell+2m}{4G_4}\, = \frac{ \bar{R}_{x_4}  \bar{R}_{x_5}  \bar{R}_y\, ( \bar{\ell}+2 \bar{m})}{4\pi^3 l_s^4 } \label{eq:Mass}\\
& \,=\,\frac{\bar{R}_{x_4}  \bar{R}_{y}  \left(\bar{R}_{x_5} \sqrt{4  \bar{R}_{x_5}^2+ \bar{R}_y^2} +\sqrt{4 N^2+ \bar{R}_{x_5}^2  \bar{R}_y^2}\right) }{16 \pi^3 l_s^4}\,.\nonumber
\end{align}

This mass formula is very illustrative. First,  we can see that the mass remains finite when $N=0$. So the uplifton can be thought of as a topological soliton of mass
\begin{equation}
\mathcal{M}_\text{topo} = \frac{\bar{R}_{x_4}  \bar{R}_{y}  \bar{R}_{x_5} \left(\sqrt{4  \bar{R}_{x_5}^2+ \bar{R}_y^2} + \bar{R}_y\right) }{16 \pi^3 l_s^4}\,,
\end{equation}
to which one adds fluxes corresponding to D5 charges. Remembering the factors of $g_s$ in the definition of $\bar{R}_{x_4}, \bar{R}_{x_5}$ and $\bar{R}_y$ \eqref{rescaling}, we can see that the mass of this soliton is proportional to $g_s^{-2}$, exactly as one expects for a gravitational soliton. Furthermore, since in this soliton the $x_4$ direction is not fibered, but $x_5$ and $y$ are, the dependence of the mass on $\bar{R}_{x_4}, \bar{R}_{x_5}$ and $\bar{R}_y$ when the other radii are kept fixed is $(\bar{R}_{x_4})^1$, $(\bar{R}_{x_5})^2$ and $(\bar{R}_y)^2$, again as one expects. 

Furthermore, the second square root  of \eqref{eq:Mass} looks exactly like the mass of a bound state of a topological soliton with $2N$ objects of mass proportional to $g_s^{-1}$.  Hence, one may na\"ively conclude that the side bubbles of the uplifton are bound states of D5 branes and topological solitons. However, this is not what happens: the ADM mass of $2 N$ D5 branes wrapping $x_{1,2,3,4,5}$ is
\begin{equation}
\mathcal{M}_\text{BPS}= 2\times \frac{ q}{4 G_4} =  \frac{N \bar{R}_{x_4} \bar{R}_{x_5} }{4 \pi^3 l_s^4},
\end{equation}
while the $N \gg \bar{R}_{x_5} \bar{R}_{y} $ limit\footnote{This limit can be achieved when $ \bar{R}_{y} \ll \bar{R}_{x_5}$ and $N \lesssim \bar{R}_{x_5}^2 $.} of the second square root in \eqref{eq:Mass}  gives a mass contribution proportional  to $ N \bar{R}_{x_4} \bar{R}_{y}$ instead,
\begin{equation}
\mathcal{M}_\text{flux}=   \frac{N \bar{R}_{x_4} \bar{R}_{y} }{8 \pi^3 l_s^4},
\end{equation}

Given that in this regime of parameters the growth of the soliton mass with $N$ is linear, one may ask whether it is possible to have an uplifton with  $ \mathcal{M}_\text{flux} > \mathcal{M}_\text{BPS} $, which could lower its energy by tunneling emission of a D5 and a $\overline{\text{D5}}$  brane.  Using Equation \eqref{constraint},  this does not happen, neither in the regime where the mass of the uplifton grows linearly with $N$, nor in any other regime of parameters.

Since our purpose is to use the uplifton to uplift the cosmological constant of a flux compactification in which supergravity can be trusted, the parameters  $\bar{R}_{y}, \bar{R}_{x_4}$ and $ \bar{R}_{x_5}$ have to be large (they correspond to the extra-dimension sizes in units of $\sqrt{g_s} l_s$).  One can check that in this regime of parameters the mass of the lightest uplifton (with $N=1$) is necessarily heavier than the mass of two BPS D5 branes. However, as $N$ increases, the mass of the uplifton can become smaller than the mass of $2N$ D5 branes. This happens because the binding energy of the brane and antibrane regions becomes of the same order as the energy of the branes

\vspace*{-.5cm}
\section{Discussion}
\label{sec:Disc}
\vspace*{-.2cm}

We have constructed the simplest example of an uplifton: a smooth solution that has three topologically-nontrivial cycles: a neutral one in the middle and two external ones with fluxes corresponding to  D5 and $\overline{\text{D5}}$ charges. The dependence of the uplifton mass on the charges and the size of the compact directions is exactly what one expects from a topologically-nontrivial solution with fluxes.

It is remarkable that our uplifton has exactly the same structure as the solution one might expect from the geometrical transition of D5 and $\overline{\text{D5}}$ branes studied in \cite{Aganagic:2006ex} (depicted in Figure 2 in that paper): The two-cycle wrapped by the branes shrinks at the two locations of the branes, giving rise to a flux-less topologically-nontrivial three cycle between the branes. Furthermore, the three-cycles with positive and negative flux that, before the geometric transition, were shrinking at the position of the five-branes, now become large. Hence, the configuration of  \cite{Aganagic:2006ex}  should backreact in a three-bubble solution, with a neutral bubble in the middle and two equal and oppositely-charged ones on the sides. The only difference is that in our uplifton the three-cycles have an $S^1$ that is trivially fibered over an $S^2$, while in a more general solution one may expect a more exotic fibration.

As we mentioned in the introduction, the smoothness and neutrality of the uplifton make it a more controlled uplift ingredient than $\overline{\text{D3}}$ branes. However, in the regime of parameters where we have supergravity control ($ \bar{R}_{y}, \bar{R}_{x_4}, \bar{R}_{x_5} > 1$) the uplifton is heavier than $\overline{\text{D3}}$ branes. Hence,  in order to use it for uplifting one has to place it in a high-warp region of the compactification manifold. It would be intersting to establish whether this can be done using for example the Klabanov-Strassler throat \cite{Klebanov:2000hb}.

The most important question that our analysis does not answer is whether the uplifton is perturbatively stable. The Kaluza-Klein bubbles that compose the uplifton are known to be unstable in vacuum \cite{Witten:1981gj}, so this is a nontrivial possibility.  However, as shown in \cite{Bah:2021irr,Stotyn:2011tv,Miyamoto:2006nd}, when these bubbles are wrapped by electromagnetic flux this instability can disappear. Furthermore, in the absence of fluxes it is possible that our bubbles can annihilate each other as it can happen for bound states of black holes and bubbles in vacuum \cite{Elvang:2002br}. It would be very interesting to explore these possibilities in future projects.

\noindent
{\bf Acknowledgments} We would like to thank Ibou Bah, Mariana Gra\~na and Severin L\"ust for useful discussions. This work was supported in part by the ERC Grants 772408 ``Stringlandscape'' and 787320 ``QBH Structure'', and by the NSF grant PHY-2112699.

\bibliography{microstates}

\begin{thebibliography}{20}
\expandafter\ifx\csname natexlab\endcsname\relax\def\natexlab#1{#1}\fi
\expandafter\ifx\csname bibnamefont\endcsname\relax
  \def\bibnamefont#1{#1}\fi
\expandafter\ifx\csname bibfnamefont\endcsname\relax
  \def\bibfnamefont#1{#1}\fi
\expandafter\ifx\csname citenamefont\endcsname\relax
  \def\citenamefont#1{#1}\fi
\expandafter\ifx\csname url\endcsname\relax
  \def\url#1{\texttt{#1}}\fi
\expandafter\ifx\csname urlprefix\endcsname\relax\def\urlprefix{URL }\fi
\providecommand{\bibinfo}[2]{#2}
\providecommand{\eprint}[2][]{\url{#2}}

\bibitem[{\citenamefont{Maldacena and Nunez}(2001)}]{Maldacena:2000mw}
\bibinfo{author}{\bibfnamefont{J.~M.} \bibnamefont{Maldacena}}
  \bibnamefont{and} \bibinfo{author}{\bibfnamefont{C.}~\bibnamefont{Nunez}},
  \bibinfo{journal}{Int. J. Mod. Phys. A} \textbf{\bibinfo{volume}{16}},
  \bibinfo{pages}{822} (\bibinfo{year}{2001}), \eprint{hep-th/0007018}.

\bibitem[{\citenamefont{Kachru et~al.}(2003)\citenamefont{Kachru, Kallosh,
  Linde, and Trivedi}}]{Kachru:2003aw}
\bibinfo{author}{\bibfnamefont{S.}~\bibnamefont{Kachru}},
  \bibinfo{author}{\bibfnamefont{R.}~\bibnamefont{Kallosh}},
  \bibinfo{author}{\bibfnamefont{A.~D.} \bibnamefont{Linde}}, \bibnamefont{and}
  \bibinfo{author}{\bibfnamefont{S.~P.} \bibnamefont{Trivedi}},
  \bibinfo{journal}{Phys. Rev.} \textbf{\bibinfo{volume}{D68}},
  \bibinfo{pages}{046005} (\bibinfo{year}{2003}), \eprint{hep-th/0301240}.

\bibitem[{\citenamefont{Bena et~al.}(2010)\citenamefont{Bena, Grana, and
  Halmagyi}}]{Bena:2009xk}
\bibinfo{author}{\bibfnamefont{I.}~\bibnamefont{Bena}},
  \bibinfo{author}{\bibfnamefont{M.}~\bibnamefont{Grana}}, \bibnamefont{and}
  \bibinfo{author}{\bibfnamefont{N.}~\bibnamefont{Halmagyi}},
  \bibinfo{journal}{JHEP} \textbf{\bibinfo{volume}{1009}}, \bibinfo{pages}{087}
  (\bibinfo{year}{2010}), \eprint{0912.3519}.

\bibitem[{\citenamefont{Bena et~al.}(2019)\citenamefont{Bena, Dudas, Gra\~na,
  and L\"ust}}]{Bena:2018fqc}
\bibinfo{author}{\bibfnamefont{I.}~\bibnamefont{Bena}},
  \bibinfo{author}{\bibfnamefont{E.}~\bibnamefont{Dudas}},
  \bibinfo{author}{\bibfnamefont{M.}~\bibnamefont{Gra\~na}}, \bibnamefont{and}
  \bibinfo{author}{\bibfnamefont{S.}~\bibnamefont{L\"ust}},
  \bibinfo{journal}{Fortsch. Phys.} \textbf{\bibinfo{volume}{67}},
  \bibinfo{pages}{1800100} (\bibinfo{year}{2019}), \eprint{1809.06861}.

\bibitem[{\citenamefont{Bena et~al.}(2014)\citenamefont{Bena, Grana,
  Kuperstein, and Massai}}]{Bena:2014jaa}
\bibinfo{author}{\bibfnamefont{I.}~\bibnamefont{Bena}},
  \bibinfo{author}{\bibfnamefont{M.}~\bibnamefont{Grana}},
  \bibinfo{author}{\bibfnamefont{S.}~\bibnamefont{Kuperstein}},
  \bibnamefont{and} \bibinfo{author}{\bibfnamefont{S.}~\bibnamefont{Massai}}
  (\bibinfo{year}{2014}), \eprint{1410.7776}.

\bibitem[{\citenamefont{Bena et~al.}(2022)\citenamefont{Bena, Dudas, Gra\~na,
  Lo~Monaco, and Toulikas}}]{Bena:2022ive}
\bibinfo{author}{\bibfnamefont{I.}~\bibnamefont{Bena}},
  \bibinfo{author}{\bibfnamefont{E.}~\bibnamefont{Dudas}},
  \bibinfo{author}{\bibfnamefont{M.}~\bibnamefont{Gra\~na}},
  \bibinfo{author}{\bibfnamefont{G.}~\bibnamefont{Lo~Monaco}},
  \bibnamefont{and} \bibinfo{author}{\bibfnamefont{D.}~\bibnamefont{Toulikas}}
  (\bibinfo{year}{2022}), \eprint{2211.14381}.

\bibitem[{\citenamefont{Aharony et~al.}(2005)\citenamefont{Aharony, Antebi, and
  Berkooz}}]{Aharony:2005ez}
\bibinfo{author}{\bibfnamefont{O.}~\bibnamefont{Aharony}},
  \bibinfo{author}{\bibfnamefont{Y.~E.} \bibnamefont{Antebi}},
  \bibnamefont{and} \bibinfo{author}{\bibfnamefont{M.}~\bibnamefont{Berkooz}},
  \bibinfo{journal}{Phys. Rev. D} \textbf{\bibinfo{volume}{72}},
  \bibinfo{pages}{106009} (\bibinfo{year}{2005}), \eprint{hep-th/0508080}.

\bibitem[{\citenamefont{Bah and Heidmann}(2021{\natexlab{a}})}]{Bah:2020pdz}
\bibinfo{author}{\bibfnamefont{I.}~\bibnamefont{Bah}} \bibnamefont{and}
  \bibinfo{author}{\bibfnamefont{P.}~\bibnamefont{Heidmann}},
  \bibinfo{journal}{JHEP} \textbf{\bibinfo{volume}{09}}, \bibinfo{pages}{147}
  (\bibinfo{year}{2021}{\natexlab{a}}), \eprint{2012.13407}.

\bibitem[{\citenamefont{Bah and Heidmann}(2021{\natexlab{b}})}]{Bah:2021owp}
\bibinfo{author}{\bibfnamefont{I.}~\bibnamefont{Bah}} \bibnamefont{and}
  \bibinfo{author}{\bibfnamefont{P.}~\bibnamefont{Heidmann}},
  \bibinfo{journal}{JHEP} \textbf{\bibinfo{volume}{09}}, \bibinfo{pages}{128}
  (\bibinfo{year}{2021}{\natexlab{b}}), \eprint{2106.05118}.

\bibitem[{\citenamefont{Bah and Heidmann}(2021{\natexlab{c}})}]{Bah:2021rki}
\bibinfo{author}{\bibfnamefont{I.}~\bibnamefont{Bah}} \bibnamefont{and}
  \bibinfo{author}{\bibfnamefont{P.}~\bibnamefont{Heidmann}},
  \bibinfo{journal}{JHEP} \textbf{\bibinfo{volume}{10}}, \bibinfo{pages}{165}
  (\bibinfo{year}{2021}{\natexlab{c}}), \eprint{2107.13551}.

\bibitem[{\citenamefont{Heidmann}(2022)}]{Heidmann:2021cms}
\bibinfo{author}{\bibfnamefont{P.}~\bibnamefont{Heidmann}},
  \bibinfo{journal}{JHEP} \textbf{\bibinfo{volume}{02}}, \bibinfo{pages}{162}
  (\bibinfo{year}{2022}), \eprint{2112.03279}.

\bibitem[{\citenamefont{Bah et~al.}(2022)\citenamefont{Bah, Heidmann, and
  Weck}}]{Bah:2022yji}
\bibinfo{author}{\bibfnamefont{I.}~\bibnamefont{Bah}},
  \bibinfo{author}{\bibfnamefont{P.}~\bibnamefont{Heidmann}}, \bibnamefont{and}
  \bibinfo{author}{\bibfnamefont{P.}~\bibnamefont{Weck}},
  \bibinfo{journal}{JHEP} \textbf{\bibinfo{volume}{08}}, \bibinfo{pages}{269}
  (\bibinfo{year}{2022}), \eprint{2203.12625}.

\bibitem[{\citenamefont{Bah and Heidmann}(2022)}]{Bah:2022pdn}
\bibinfo{author}{\bibfnamefont{I.}~\bibnamefont{Bah}} \bibnamefont{and}
  \bibinfo{author}{\bibfnamefont{P.}~\bibnamefont{Heidmann}}
  (\bibinfo{year}{2022}), \eprint{2210.06483}.

\bibitem[{\citenamefont{Aganagic et~al.}(2008)\citenamefont{Aganagic, Beem,
  Seo, and Vafa}}]{Aganagic:2006ex}
\bibinfo{author}{\bibfnamefont{M.}~\bibnamefont{Aganagic}},
  \bibinfo{author}{\bibfnamefont{C.}~\bibnamefont{Beem}},
  \bibinfo{author}{\bibfnamefont{J.}~\bibnamefont{Seo}}, \bibnamefont{and}
  \bibinfo{author}{\bibfnamefont{C.}~\bibnamefont{Vafa}},
  \bibinfo{journal}{Nucl. Phys. B} \textbf{\bibinfo{volume}{789}},
  \bibinfo{pages}{382} (\bibinfo{year}{2008}), \eprint{hep-th/0610249}.

\bibitem[{\citenamefont{Klebanov and Strassler}(2000)}]{Klebanov:2000hb}
\bibinfo{author}{\bibfnamefont{I.~R.} \bibnamefont{Klebanov}} \bibnamefont{and}
  \bibinfo{author}{\bibfnamefont{M.~J.} \bibnamefont{Strassler}},
  \bibinfo{journal}{JHEP} \textbf{\bibinfo{volume}{0008}}, \bibinfo{pages}{052}
  (\bibinfo{year}{2000}), \eprint{hep-th/0007191}.

\bibitem[{\citenamefont{Witten}(1982)}]{Witten:1981gj}
\bibinfo{author}{\bibfnamefont{E.}~\bibnamefont{Witten}},
  \bibinfo{journal}{Nucl. Phys. B} \textbf{\bibinfo{volume}{195}},
  \bibinfo{pages}{481} (\bibinfo{year}{1982}).

\bibitem[{\citenamefont{Bah et~al.}(2021)\citenamefont{Bah, Dey, and
  Heidmann}}]{Bah:2021irr}
\bibinfo{author}{\bibfnamefont{I.}~\bibnamefont{Bah}},
  \bibinfo{author}{\bibfnamefont{A.}~\bibnamefont{Dey}}, \bibnamefont{and}
  \bibinfo{author}{\bibfnamefont{P.}~\bibnamefont{Heidmann}}
  (\bibinfo{year}{2021}), \eprint{2112.11474}.

\bibitem[{\citenamefont{Stotyn and Mann}(2011)}]{Stotyn:2011tv}
\bibinfo{author}{\bibfnamefont{S.}~\bibnamefont{Stotyn}} \bibnamefont{and}
  \bibinfo{author}{\bibfnamefont{R.~B.} \bibnamefont{Mann}},
  \bibinfo{journal}{Phys. Lett. B} \textbf{\bibinfo{volume}{705}},
  \bibinfo{pages}{269} (\bibinfo{year}{2011}), \eprint{1105.1854}.

\bibitem[{\citenamefont{Miyamoto and Kudoh}(2006)}]{Miyamoto:2006nd}
\bibinfo{author}{\bibfnamefont{U.}~\bibnamefont{Miyamoto}} \bibnamefont{and}
  \bibinfo{author}{\bibfnamefont{H.}~\bibnamefont{Kudoh}},
  \bibinfo{journal}{JHEP} \textbf{\bibinfo{volume}{12}}, \bibinfo{pages}{048}
  (\bibinfo{year}{2006}), \eprint{gr-qc/0609046}.

\bibitem[{\citenamefont{Elvang and Horowitz}(2003)}]{Elvang:2002br}
\bibinfo{author}{\bibfnamefont{H.}~\bibnamefont{Elvang}} \bibnamefont{and}
  \bibinfo{author}{\bibfnamefont{G.~T.} \bibnamefont{Horowitz}},
  \bibinfo{journal}{Phys. Rev. D} \textbf{\bibinfo{volume}{67}},
  \bibinfo{pages}{044015} (\bibinfo{year}{2003}), \eprint{hep-th/0210303}.

\end{thebibliography}

\end{document}